\documentclass[10pt,pra,aps,twocolumn,superscriptaddress,floatfix,notitlepage,nofootinbib]{revtex4-1}
\usepackage[latin1]{inputenc}
\usepackage{natbib}
\usepackage{amsmath}
\usepackage{amsfonts}
\usepackage{amssymb}
\usepackage{bm}

\usepackage[pdftex]{graphicx}

\usepackage{graphicx}
\usepackage[squaren,cdot]{SIunits}
\usepackage{color}

\newcommand\e{\mathrm e}
\newcommand\kB{k_\mathrm B}
\newcommand\LL{\mathrm L}
\newcommand\RR{\mathrm R}
\newcommand\HH{\mathrm H}
\newcommand\CC{\mathrm C}
\newcommand\T{\mathcal T}
\newcommand\STOP{\mathrm{stop}}

\begin{document}
	\title{Autonomous quantum absorption refrigerators}
	\author{Sreenath K. Manikandan}
	\email{skizhakk@ur.rochester.edu}
	\affiliation{Department of Physics and Astronomy, University of Rochester, Rochester, NY 14627, USA}
	\affiliation{Center for Coherence and Quantum Optics, University of Rochester, Rochester, NY 14627, USA}
		\author{\'Etienne Jussiau}
		\email{ejussiau@ur.rochester.edu}
	\affiliation{Department of Physics and Astronomy, University of Rochester, Rochester, NY 14627, USA}
	\affiliation{Center for Coherence and Quantum Optics, University of Rochester, Rochester, NY 14627, USA}
	\author{Andrew N. Jordan}
	\email{jordan@pas.rochester.edu}
	\affiliation{Department of Physics and Astronomy, University of Rochester, Rochester, NY 14627, USA}
	\affiliation{Center for Coherence and Quantum Optics, University of Rochester, Rochester, NY 14627, USA}    
	\affiliation{Institute for Quantum Studies, Chapman University, Orange, CA, 92866, USA}
\date{\today}
	\begin{abstract}
We propose a quantum absorption refrigerator using the quantum physics of resonant tunneling through quantum dots. The cold and hot reservoirs are fermionic leads, tunnel coupled via quantum dots to a central fermionic cavity, and we propose configurations in which the heat absorbed from the (very hot) central cavity is used as a resource to selectively transfer heat from the cold reservoir on the left, to the hot reservoir on the right. The heat transport in the device is particle---hole symmetric; we find two regimes of cooling as a function of the energy of the dots---symmetric with respect to the Fermi energy of the reservoirs---and we associate them to heat transfer by electrons above the Fermi level, and holes below the Fermi level, respectively. We also discuss optimizing the cooling effect by fine-tuning the energy of the dots as well as their linewidth, and characterize regimes where the transport is thermodynamically reversible such that Carnot Coefficent of Performance is achieved with zero cooling power delivered.
	\end{abstract}
	\maketitle
\section{Introduction}
 
Converting otherwise wasted heat to perform useful work in the nanoscale is an open problem that spans across almost all disciplines of applied science~\cite{costache2010experimental,hwang2013proposal,sothmann2012rectification,serreli2007molecular,sothmann2014thermoelectric,kosloff2014quantum}, including computing~\cite{pekola2015towards,senior2020heat,landauer1982uncertainty,berut2012experimental,keyes1970minimal,landauer1961irreversibility}, where there is a lower bound on dissipated heat per cycle of irreversible computation, given by the Landauer's bound.\footnote{Landauer's bound is a lower bound which predicts that the minimum heat generated in erasing a classical bit worth of information is at least $\kB T\log{2}$. The bound can be saturated by implementing erasure as a thermodynamic cycle consisting of adiabatic and isothermal processes, where $T$ is the temperature of the heat bath facilitating the isotherms~\cite{landauer1961irreversibility}.}  Managing the excess heat generated in circuits is also crucial for various quantum computing platforms currently available, such as superconducting qubits and matter based spin qubits, where cooling down to sub-kelvin temperatures is a must~\cite{pekola2015towards}. Besides, efficient cooling below $\unit4\kelvin$ is a necessity for enhancing the performance of radiation detectors and charge sensors, with benefits also extending to various medical applications, including magnetic resonant imaging~\cite{walker1989miniature,giazotto2006opportunities,pobell2007matter}. Furthermore, sub-kelvin cooling is essential for exploiting quantum physics in the mesoscopic regime for quantum device applications, and for  nano-scale energy harvesting with quantum dots~\cite{sothmann2014thermoelectric}. The increasing demand for achieving temperatures nearing absolute zero is largely fulfilled by state of the art dilution refrigerators, which can achieve base temperature down to about $\unit{10}{\milli\kelvin}$; even then, localized dissipation of heat remains a major issue to be addressed in places including quantum circuits, where it is a limiting factor for achieving coherent, non-local manipulation of quantum information in various quantum computing platforms currently available~\cite{ioffe2004decoherence,cywinski2009electron,mohseni2003experimental,qiao2020conditional,das2009decoherence,pekola2015towards,savin2006thermal,vandersypen2017interfacing,labaziewicz2008suppression}.

Cooling has always been an exciting problem in thermodynamics~\cite{pobell2007matter}, and the advent of quantum technologies presented more recent opportunities for novel refrigeration schemes which can be integrated into various computing platforms, and further cool down the devices below ambient temperatures~\cite{giazotto2006opportunities}. Some examples of such cooling techniques in solid state include nuclear demagnetization~\cite{pobell2007matter}, voltage biased junction refrigerators~\cite{leivo1996efficient,pekola1999nis}, and Josephson junction based refrigerators~\cite{solinas2016microwave,giazotto2012josephson}. Also see Ref.~\cite{giazotto2006opportunities} and references therein. Solid state refrigeration schemes using adiabatic magnetization of a superconductor have also been proposed, which is particularly useful as a cooling mechanism below the superconducting critical temperatures~\cite{manikandan2019superconducting,dolcini2009adiabatic,svidzinsky2002possible}.

An alternate approach to dealing with excess heat in quantum circuits is to recycle this heat as a resource to power other quantum thermal machines, such as quantum heat engines and quantum refrigerators~\cite{kosloff2014quantum,campisi2014fluctuation,whitney2014most,quan2007quantum,sothmann2014thermoelectric,jordan2013powerful,sanchez2013correlations,sothmann2013powerful,sothmann2012rectification,benenti2017fundamental}. A refrigerator powered by a dissipative heat source is conventionally called an absorption refrigerator~\cite{srikhirin2001review,herold2016absorption}; The principles of such a cooling technique were known since the 1700s, and further developed through the early twentieth century, as an alternative to the standard compression based refrigerators~\cite{gordon2000cool,albert1930refrigeration,gosney1982principles}. Albeit having lower coefficient of performance, the utility of absorption refrigerators emerges from their unique approach to cooling, where excess heat, potentially from a dissipative heat source, is used as a resource to run the cooling cycle. In an evaporation based absorption refrigeration cycle, the evaporation of a cooling agent at the cold reservoir generates the cooling power. A low vapor pressure required for evaporation is maintained by an absorbing fluid, which reduces the vapor pressure of the cooling agent by absorbing it in the vapor phase. Subsequently, the absorbing agent is heated by an external heat source, which releases the cooling agent, now hotter than its ambient temperature. The excess heat is released into the hot reservoir (typically the environment), and the cooling agent condenses as it flows back into the cold reservoir. The cycle repeats.

Absorption refrigerators that operate at mesoscopic scales, where quantum effects are relevant, have also emerged as one among the prototypical systems to probe thermodynamics in the quantum regime~\cite{benenti2017fundamental,levy2012quantum,maslennikov2019quantum,brask2015small,mitchison2016realising,maslennikov2019quantum,jaliel2019experimental}. A canonical model would consist of three reservoirs---cold ($\LL$), hot ($\RR$),  and hotter ($\HH$), where $T_\LL\leq T_\RR\leq T_\HH$---and quantum systems (possibly qubits) interacting among themselves, as well as with the reservoirs. The interactions are such that the spontaneous flow of heat from $\HH\to\RR$ also induces a flow of heat from $\LL\to\RR$, resulting in further cooling of the reservoir L. See for instance, Ref.~\cite{erdman2018absorption} where a proposal for such an absorption refrigerator using Coulomb-coupled quantum dots/metallic islands is discussed. It has also been pointed out that quantum coherent effects may enhance the performance of absorption refrigerators in the quantum regime~\cite{correa2014quantum,mitchison2015coherence}, suggesting that absorption refrigerators may be used to probe quantum advantages in the operation of thermal machines. Thermoelectric effects in nanoscale devices are tighly linked to their energy-filtering properties~\cite{mahan1996best}. Allowing a flow of electrons between two terminals at certain energies only can give rise to a flow of charge against an electrochemical potential or a flow of heat against a temperature bias. A thermoelectric device is then characterized by its transmission function~$\mathcal T(E)$ describing the probability for an electron at energy~$E$ to traverse the system. Typically, thermoelectric effects arise in devices for which $\mathcal T(E)$ behaves differently above and below the Fermi energy~\cite{benenti2017fundamental}. We then understand the importance of working with devices with prominent and well-characterized energy-filtering properties. This is why many experimental realizations of nanoscale thermoelectrics make use of quantum dots whose transmission function is given by a Lorentzian function centered at the resonant dot energy~\cite{prance2009electronic,hartmann2015voltage,thierschmann2015three,jaliel2019experimental}.

\begin{figure}
    \includegraphics[width=\linewidth]{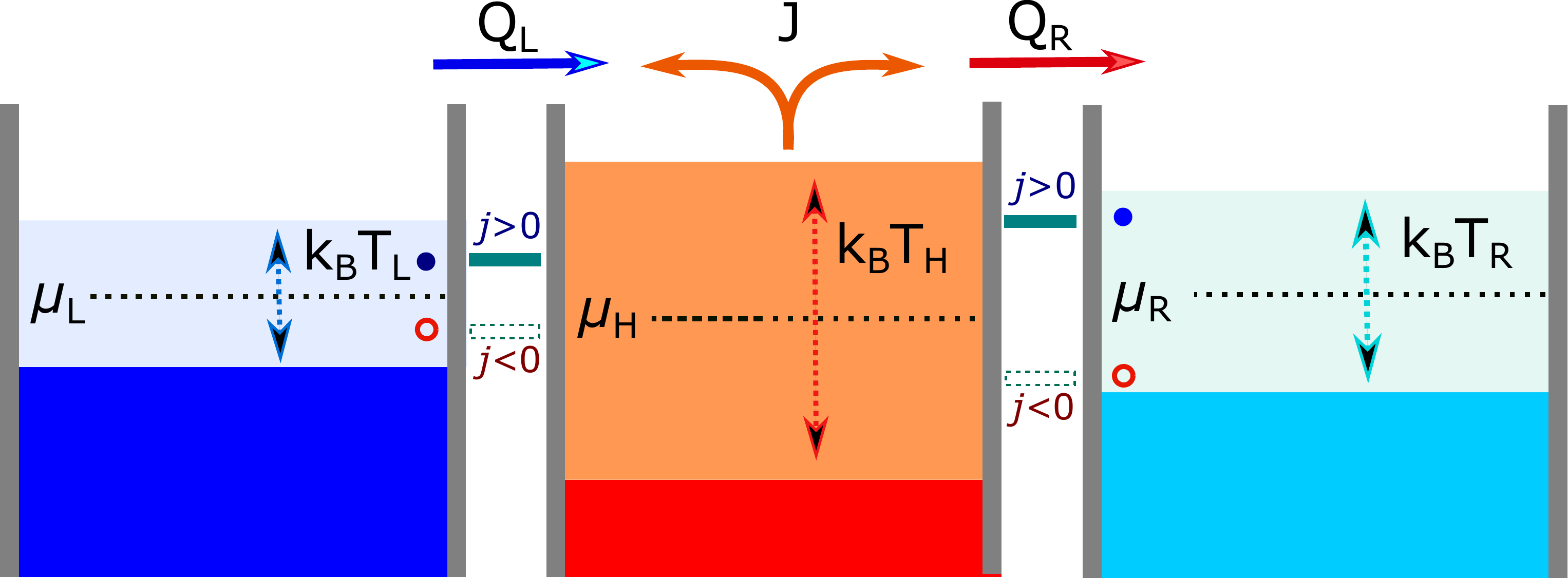}
    \caption{Schematic of particle currents ($j$) and heat currents ($Q_\LL,~Q_\RR$ and $J$) in the absorption refrigerator. The bias of the dot energies~$E_\LL$ and~$E_\RR$ w.r.t. the Fermi energy of the leads chooses whether the transport is mediated by electrons (above the Fermi energy of the leads, $j>0$), or holes (below the Fermi energy of the leads, $j<0$). Nevertheless, the heat currents are invariant under this choice of bias, depicting particle-hole symmetry in the transport problem.}
    \label{fig1}
\end{figure}

In the present article, we propose a quantum absorption refrigerator where we take advantage of the quantum physics of resonant tunneling through quantum dots to achieve the unidirectional flow of heat required for refrigeration. 
In our proposal, we consider an energy filtering configuration for the dot enegies, similar to the one considered in Refs.~\cite{jordan2013powerful,sothmann2013powerful}.  We present an experimentally viable design motivated from a recently realized  energy harvesting quantum device discussed in Ref.~\cite{jaliel2019experimental}. 

The configuration we consider is sketched in Fig.~\ref{fig1}. The reservoir~$\HH$ is a fermionic cavity, that is coupled to a cold reservoir~$\LL$ on the left, and a hot reservoir~$\RR$ on the right via quantum dots having prescribed energies. Refrigerator configurations in such architectures have been investigated in Refs.~\cite{prance2009electronic,wang2018nonlinear}, where the goal is to cool down the central cavity, H. A variant of this where quantum dots are replaced by superlattices is presented in Ref.~\cite{choi2015three}. Similar tunnel coupling to cool down a central metallic reservoir using selective transfer of hot electrons and holes to left and right reservoirs has also been proposed~\cite{edwards1995cryogenic}.   In contrast, our present study investigates whether it is possible to think of the fermionic cavity as a hot spot in a circuit, that allows to extract finite cooling power from the cold reservoir L. In addition to that, we also provide a complete thermodynamic characterization of the device and discuss its optimal, and stopping configurations. Our analysis also makes simple connections to the particle-hole symmetry in the transport problem from a thermodynamic point of view,  which reveals two equivalent bias configurations for the operation of our absorption refrigerator. They correspond to energy of the dots positioned above the Fermi level (mediated by hot electrons), and below the Fermi level (mediated by hot holes) respectively. 

Note that a pedagogical analogy can be made to a conventional refrigerator context where work has to be supplied to refrigerate reservoir~$\LL$ relative to reservoir~$\RR$. Here, one can think of the reservoirs~$\HH$ and~$\RR$ as analogous to the hot and cold reservoirs of an engine which supply the necessary work required for cooling in a conventional refrigerator, where the heat flow from $\HH\to\RR$ generates the fiducial work to run the refrigeration of the reservoir~$\LL$ relative to  the reservoir~$\RR$.

In the discussions which follow, we provide a systematic characterization of our quantum absorption refrigerator in its steady state. The system obeys both particle and energy current conservation laws in the steady state. We further assume that the chemical potentials are identical for the reservoirs~$\LL$ and~$\RR$, so as to assure that our device qualifies as an absorption refrigerator. Indeed, in such a situation, the refrigeration is solely powered by the heat provided by the hot cavity~$\HH$, contrary to standard nanoscale refrigerators where a voltage bias enables electric power generation to fuel the refrigerator~\cite{pekola1999nis,giazotto2006opportunities}. From the point of view of electrons leaving the cold reservoir, they have to gain definite energy from the cavity~$\HH$ to overcome the temperature difference and exit to the hot reservoir on the right. However, we can add a voltage bias to our setup to design hybrid devices that use the heat from the cavity both to cool down reservoir~$\LL$ and generate electric power in reservoir~$\RR$ (if $\mu_\LL<\mu_\RR$), or, conversely, we can imagine a ``doped'' absorption refrigerator whose performance is improved using electric power alongside the heat from the cavity (if $\mu_\LL>\mu_\RR$).

A crucial assumption we make is that the hot and cold reservoirs are connected to some external circuit, while the cavity is in thermal equilibrium with a separate heat reservoir. As such, reservoirs~$\LL$ and~$\RR$ can exchange particles with their environment, and their chemical potentials~$\mu_\LL$ and~$\mu_\RR$ can then be imposed externally. On the contrary, the number of particles in the cavity is constant and its chemical potential is then fixed by particle conservation.\footnote{This is in the same spirit as the distinction between the canonical and grand canonical ensemble in statistical mechanics.} Furthermore, we assume that strong inelastic electron-electron or electron-phonon interactions taking place in the cavity cause electrons entering it to relax on a time scale much shorter than the time they will spend there. As such, electron populations in the cavity are described by the usual Fermi factors, where the chemical potential is determined through particle conservation.

This article is organized as follows. We first discuss our model in detail, in light of the conservation laws. We then extend this discussion to characterize the laws of thermodynamics for our absorption refrigerator, assuming vanishing linewidth for the quantum dots. In subsequent sections we prescribe methods to optimize cooling power over energy of the dots, as well as temperature of the leads involved. We also numerically investigate optimizing the cooling power over the finite linewidth of the quantum dots, and present estimates of experimentally achievable figures of merit  for our quantum absorption refrigerator.

\begin{figure}
    \includegraphics[width=\linewidth]{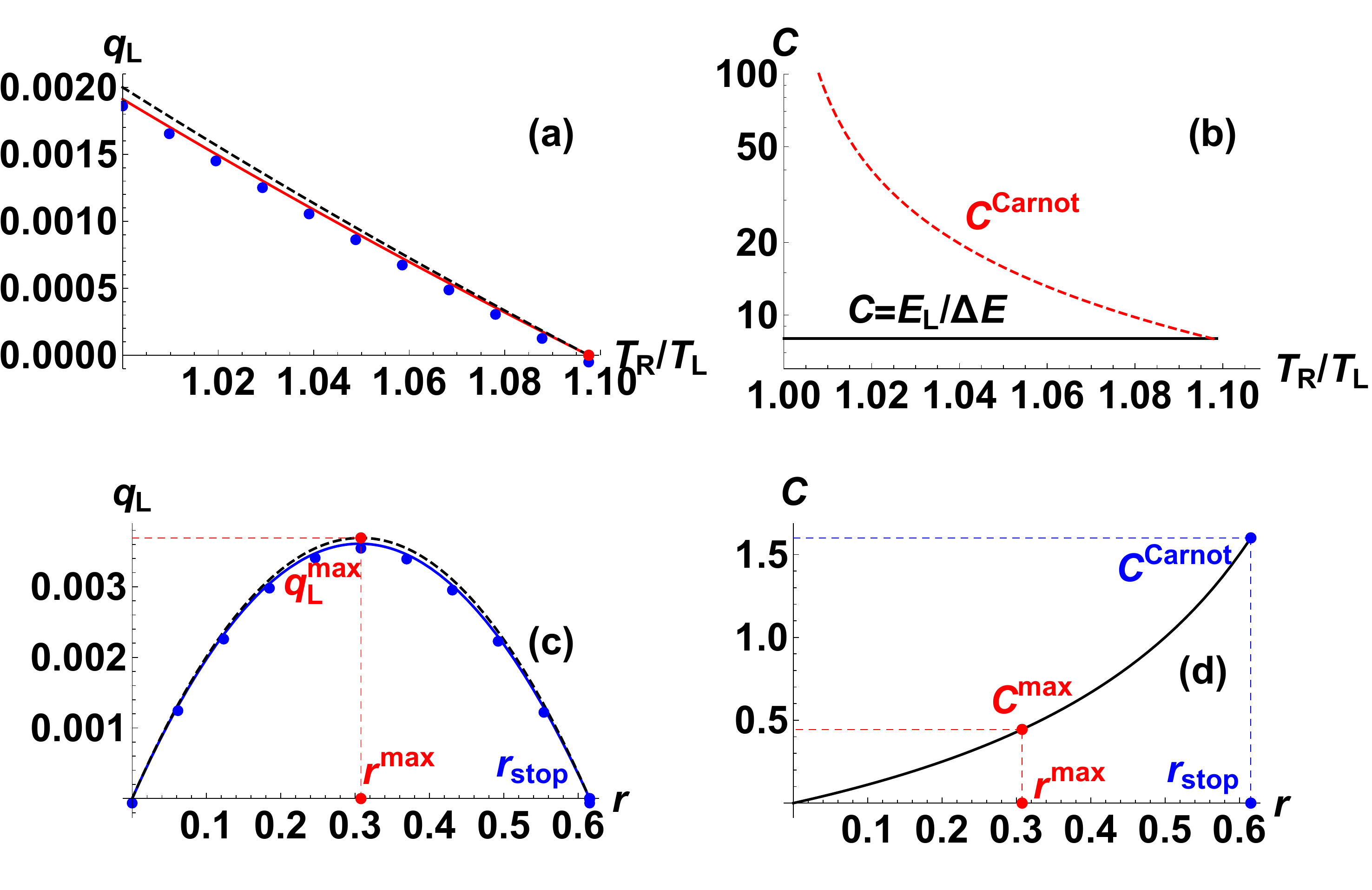}
    \caption{(a) The cooling power $Q_\LL$ in dimensionless units $q_\LL=\hbar Q_\LL/(\gamma\kB T_\LL)$, as a function of $T_\RR/T_\LL$. We also compare numerical calculation for $q_\LL$ for small, but finite, linewidth (blue dots) with exact predictions assuming $\delta$ transmission function (joined, red), as well as linear response regime results (dashed, black).  We choose $\mu_\LL=\mu_\RR=0$, $T_\HH=5T_\LL$, $E_\LL=0.4\kB T_\LL$, $E_\RR=0.45\kB T_\LL$, $\gamma=10^{-4} E_\RR$. (b) Coefficient of performance as a function of $T_\RR/T_\LL$. Note that Carnot coefficient of performance is reached at $T_\RR=T_\STOP$. (c) Comparing numerical calculation for $q_\LL$ for small, but finite, linewidth (blue dots) with exact predictions assuming $\delta$ transmission function (joined, blue), as well as linear response regime results (dashed, black) for different value of $r=E_\LL/E_\RR$. We choose $\mu_\LL=\mu_\RR=0$, $T_\RR=1.5T_\LL$, $T_\HH=5T_\RR$, $E_\RR=0.6\kB T_\LL$, $\gamma=10^{-4}E_\RR$. (d) Coefficient of performance (C) of the absorption refrigerator. It is shown that the refrigerator achieves Carnot coefficient of performance at the stopping energy, $E_\LL=r_{\STOP}E_\RR$, and that the coefficient of performance at $E_\LL^{\mathrm{max}}$ is $C^{\mathrm{max}}=(T_\RR^{-1}-T_\HH^{-1})/(2T_\LL^{-1}-T_\RR^{-1}-T_\HH^{-1})$.}
   \label{figmm}
\end{figure}
\section{The model and conservation laws}

We consider two fermionic reservoirs~$\LL$ and~$\RR$ connected via two quantum dots at energies~$E_\LL$ and~$E_\RR$ to a cavity~$\HH$ in the middle (see Fig.~\ref{fig1}). The resonant tunneling quantum dots are tunnel-coupled to the reservoirs and cavity, each contact being characterized by a tunneling rate~$\gamma_\alpha$ which corresponds to the inverse lifetime of an electron on the dot, and is also referred to as the level-width for the dot. In what follows, we will assume symmetric coupling, that is, all tunnel rates are taken equal, $\gamma_\LL=\gamma_\RR=\gamma$.

Electron populations in the leads and cavity are described by the Fermi-Dirac distributions,
\begin{equation}
    f(E-\mu_\alpha,T_\alpha)=\bigg(\e^{\frac{E-\mu_\alpha}{\kB T_\alpha}}+1\bigg)^{-1},\quad\alpha=\LL,\RR,\HH.
\end{equation}
Here $\mu_\alpha$ are the chemical potential of reservoir~$\alpha$, and $\kB$ is the Boltzmann's constant. The particle and energy currents out of reservoir~$\alpha=\LL,\RR$, denoted by $j_\alpha$ and $J_\alpha$ respectively, are given by the Landauer--B\"{u}ttiker-type expressions,
\begin{align}
    &j_\alpha=\frac{2}{h}\int\mathrm dE\,\T_\alpha(E)[f(E-\mu_\alpha,T_\alpha)-f(E-\mu_\HH,T_\HH)],\label{pcurrent}\\
    &J_\alpha=\frac{2}{h}\int\mathrm dE\,E\T_\alpha(E)[f(E-\mu_\alpha,T_\alpha)-f(E-\mu_\HH,T_\HH)],\label{ecurrent}
\end{align}
where $\T_\alpha(E)$, the transmission function for dot~$\alpha$, assumes a Lorentzian shape for resonant tunneling~\cite{buttiker1988coherent,benenti2017fundamental},
\begin{equation}
    \T_\alpha(E)=\frac{\gamma^2}{(E-E_\alpha)^2+\gamma^2}.\label{transmission}
\end{equation}
We furthermore define the heat current associated to each of the leads as,
\begin{equation}
    Q_\alpha=J_\alpha-\mu_\alpha j_\alpha.
\end{equation}

The chemical potential of the cavity cannot be chosen arbitrarily; it is constrained by particle conservation across the device,
\begin{equation}
    j_\LL+j_\RR=0.\label{pcons}
\end{equation}
The above relation clearly asserts that the net particle current out of the cavity vanishes in steady state. Consequently, the total heat current~$J$ flowing out the cavity coincides with the corresponding energy current~\cite{jordan2013powerful}, and can thus be inferred from energy conservation,
\begin{equation}
    J+J_\LL+J_\RR=0.\label{econs}
\end{equation}
$J$ is identified as the heat current driving the absorption refrigerator. 

In general, the conservation laws in Eqs.~\eqref{pcons} and~\eqref{econs} cannot be solved exactly. It is however possible in the narrow-linewidth regime, $\gamma\ll\kB T_\LL,\kB T_\RR,\kB T_\HH$, where the transmission function in Eq.~\eqref{transmission} can be approximated by $\T(E)=\pi\gamma\delta(E-E_\alpha)$ owing to the delta function limit of the Lorentzian function,
\begin{equation}
    \lim_{\gamma\to0}\frac{1}{\pi}\frac{\gamma}{x^{2}+\gamma^{2}}=\delta(x).
    \label{delta}
\end{equation}
In this regime, the conservation equations become,
\begin{widetext}
\begin{align}
    &f(E_\LL-\mu_\LL,T_\LL)-f(E_\LL-\mu_\HH,T_\HH)+f(E_\RR-\mu_\RR,T_\RR)-f(E_\RR-\mu_\HH,T_\HH)=0,\label{cc}\\
    &J+\frac{\gamma E_\LL}{\hbar}    [f(E_\LL-\mu_\LL,T_\LL)-f(E_\LL-\mu_\HH,T_\HH)]+\frac{\gamma E_\RR}{\hbar}[f(E_\RR-\mu_\RR,T_\RR)-f(E_\RR-\mu_\HH,T_\HH)]=0,\label{heat}
\end{align}
\end{widetext}
reminiscent of similar equations in Ref.~\cite{jordan2013powerful}.
Above, Eq.~\eqref{cc} suggests that the parameters in the transport problem cannot be defined independently of each other as a result of current conservation. We can therefore use Eq.~\eqref{cc} to determine the chemical potential of the cavity $\mu_\HH$ as a function of other parameters, such as the energy of the dots, and chemical potential of the external leads, $\mu_\LL$ and $\mu_\RR$ (see App.~\ref{appA}). This is because these parameters---energy of the dots and the chemical potential of the external leads---are typically fixed in a given experimental setting, which can be controlled via external voltage controls.

Solving Eq.~\eqref{cc} for the chemical potential~$\mu_\HH$ yields an exact expression the particle current across the device,
\begin{equation}
\begin{aligned}[t]
    j&=\frac\gamma\hbar\big[f(E_\LL-\mu_\LL,T_\LL)-f(E_\LL-\mu_\HH,T_\HH)\big],\\
    &=\frac\gamma\hbar\big[f(E_\RR-\mu_\HH,T_\HH)-f(E_\RR-\mu_\RR,T_\RR)\big].
\end{aligned}
\end{equation}
The heat current~$J$ then straightforwardly follows: From Eq.~\eqref{heat}, we find $J=j\Delta E$, where $\Delta E=E_\RR-E_\LL$ is the energy gain between the right and left dots. Such relation is typical of the so-called tight-coupling limit where particle and energy currents are proportional to one another. Indeed, in the narrow-linewidth limit, each electron flowing from $\LL$ to $\RR$ necessarily carries a definite amount~$\Delta E$ of energy. This property no longer holds when $\gamma$ increases as the dots' energy levels widen allowing electrons with different energies to pass.

\section{Thermodynamic analysis}

\subsection{Thermodynamics of the absorption refrigerator}

We now analyze the situation where the device is used as an absorption refrigerator and characterize it thermodynamically. We want to use the heat from the hottest cavity as a resource to induce a heat current out of the coldest reservoir without electrical power involved. Here, we assume $T_\LL<T_\RR<T_\HH$ and $\mu_\LL=\mu_\RR=\mu$. Hereafter, we set the zero of energy at~$\mu$ without loss of generality. Furthermore, for the system to function as a refrigerator, the cooling power, that is the heat current out of the cold reservoir, must be positive, namely $Q_\LL>0$. For the device at stake here, the laws of thermodynamics read,
\begin{equation}
 J+Q_\LL+Q_\RR=0,\label{1stLaw}
\end{equation}
which is the statement of global conservation of energy (first law), and the second law of thermodynamics in the Clausius form~\cite{callen2006thermodynamics}:
\begin{equation}
 \frac{J}{T_\HH}+\frac{Q_\LL}{T_\LL}+\frac{Q_\RR}{T_\RR}\le0.\label{2ndLaw}
\end{equation}
The first law of thermodynamics in Eq.~\eqref{1stLaw} enables us to eliminate one the heat currents from the entropy balance in Eq.~\eqref{2ndLaw}. As such, we obtain,
\begin{equation}
    J\bigg(\frac{1}{T_\RR}-\frac{1}{T_\HH}\bigg)\ge Q_\LL\bigg(\frac{1}{T_\LL}-\frac{1}{T_\RR}\bigg)\ge0\label{2ndLaw_1}
\end{equation}
and,
\begin{equation}
    Q_\RR\bigg(\frac{1}{T_\HH}-\frac{1}{T_\RR}\bigg)\ge Q_\LL\bigg(\frac{1}{T_\LL}-\frac{1}{T_\HH}\bigg)\ge0.
\end{equation}
Our choice $T_\HH>T_\RR$ imposes $J\ge0$, and $Q_\RR\le0$. In this situation, the heat out of the cavity~$\HH$ drives a heat current from reservoir~$\LL$ to reservoir~$\RR$, enabling cooling of the former. The coefficient of performance (COP) of the absorption refrigerator is then defined as,
\begin{equation}
    C=\frac{Q_\LL}{J}.
\end{equation}
The COP is maximum when the refrigerators operates reversibly. We refer to this upper bound as the Carnot COP, and its value can be obtained from Eq.~\eqref{2ndLaw_1},
\begin{multline}
         J\bigg(\frac{1}{T_\RR}-\frac{1}{T_\HH}\bigg)\ge Q_\LL\bigg(\frac{1}{T_\LL}-\frac{1}{T_\RR}\bigg)\\
         \implies C=\frac{Q_\LL}J\le\frac{T_\RR^{-1}-T_\HH^{-1}}{T_\LL^{-1}-T_\RR^{-1}}=C_\mathrm{Carnot}.
\end{multline}

\subsection{Vanishingly small linewidth}

These thermodynamic relations can be written in terms of the microscopic details of our device in the limit of small level width, $\gamma\ll\kB T_\LL$. In this regime, we have $Q_\LL=jE_\LL$, $Q_\RR=-jE_\RR$ and $J=j\Delta E$. The second law then reads
\begin{equation}
    j\bigg(\frac{\Delta E}{T_\HH}+\frac{E_\LL}{T_\LL}-\frac{E_\RR}{T_\RR}\bigg)\le0,\label{2ndLaw_v}
\end{equation}
and the coefficient of performance is given by
\begin{equation}
    C=\frac{E_\LL}{\Delta E}.
\end{equation}
Interestingly, there are two possible choices for the relative positions of the dot energies and Fermi level such that $J\ge0$, $Q_\LL\ge0$ and $Q_\RR\le0$. In comparison to the electric current rectification case discussed in Ref.~\cite{jordan2013powerful} where the dot energies are positioned above and below the Fermi energy, here both the dot energies are positioned either above, or below the Fermi energy. They respectively correspond to $E_\RR>E_\LL>0$, where the particle current flows from $\LL$ to $\RR$ ($j>0$), or $E_\RR<E_\LL<0$, where the particle current flow from $\RR$ to $\LL$ ($j<0$). In the former case, hot electrons are taken out of the cold reservoir, while cold electrons are injected into the cold reservoir in the latter case. Alternatively, the latter case can be viewed as the transport of hot holes below the Fermi energy (taken as the zero of energy), see Fig.~\ref{fig1}. We note that this is a manifestation of particle-hole symmetry in the underlying transport problem, which is often an overlooked aspect, but has interesting consequences; in our absorption refrigerator context, the particle-hole symmetry can be exploited as an additional freedom of choice for the biasing of dot energies relative to the Fermi energy, and this freedom could be beneficial in an experimental implementation of our proposal.

\begin{figure}
    \includegraphics[width=\linewidth]{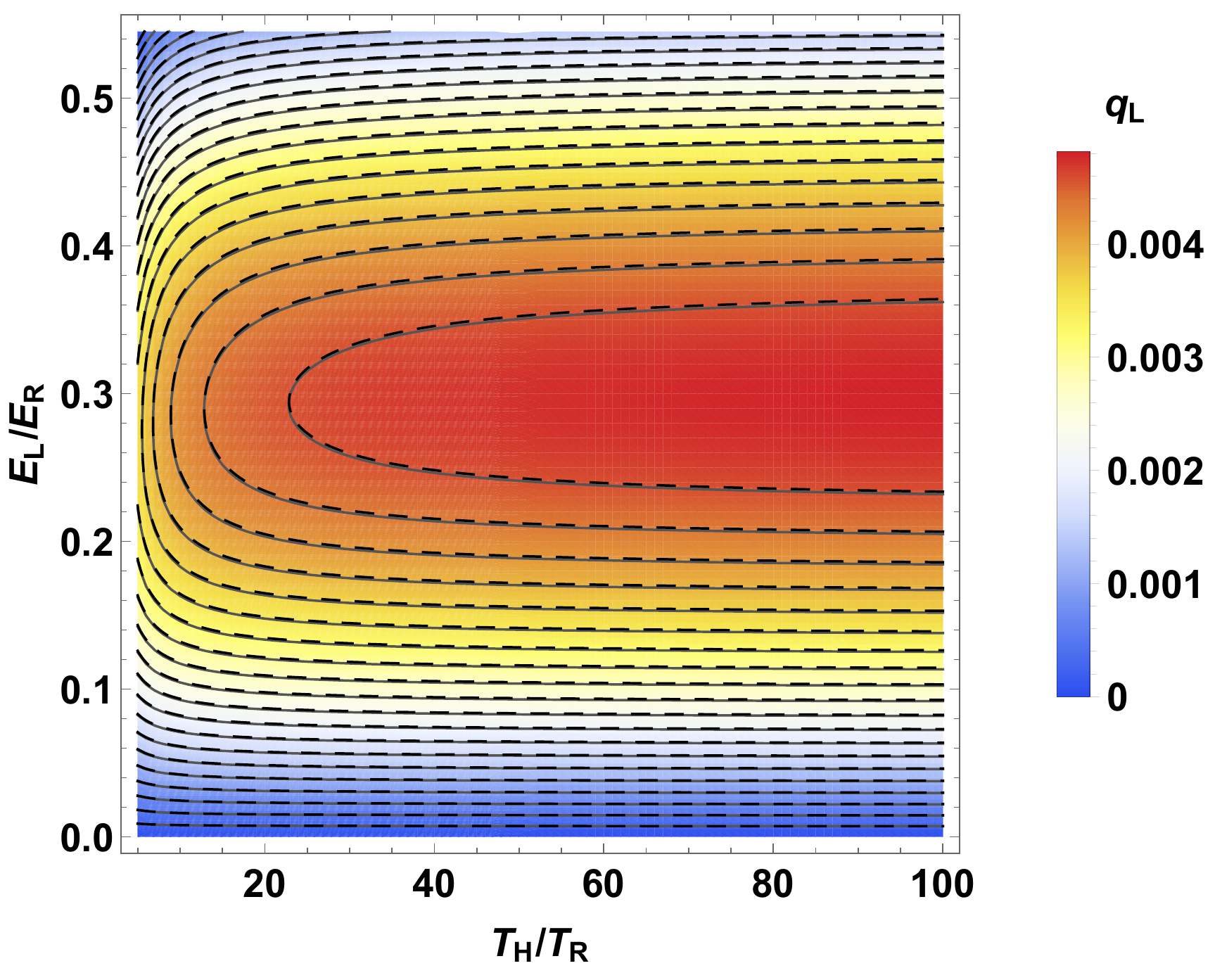}
    \caption{The cooling power~$Q_\LL$ in dimensionless units, $q_\LL=\hbar Q_\LL/(\gamma\kB T_\LL)$ as a function of dot energies, as well as temperatures. The contour lines are obtained independently in the small-linewidth limit from exact calculations (solid lines) and linear response results (dashed lines). We choose $\mu_\LL=\mu_\RR=0$, $T_\LL=0.6T_\RR$, $E_\RR=0.4\kB T_\RR$. The minimum temperature of reservoir~$\HH$ is set at $T_\HH=5T_\RR$ and the maximum value of the ratio $E_\LL/E_\RR$ is taken to be the stopping ratio~$r_\STOP$ for this minimum temperature.}
    \label{fig3}
\end{figure}

The thermodynamic analysis presented above allows us to predict the temperature of the hot reservoir~$\RR$ for which cooling power vanishes. According to Eq.~\eqref{2ndLaw_v}, for $j>0$, we must have,
\begin{equation}
    \bigg(\frac{E_\LL}{T_\LL}-\frac{E_\RR}{T_\RR}+\frac{\Delta E}{T_\HH}\bigg)\leq 0,\label{eq1}
\end{equation}
while, for $j<0$, we must have
\begin{equation}
    \bigg(\frac{E_\LL}{T_\LL}-\frac{E_\RR}{T_\RR}+\frac{\Delta E}{T_\HH}\bigg)\geq 0.\label{eq2}
\end{equation}
In either of the cases the stopping configuration corresponds to saturating the equality in Eqs.~\eqref{eq1} and~\eqref{eq2}, as the entropy change becomes zero and the system becomes thermodynamically reversible. Solving for $T_\RR=T_\STOP$, we obtain,
\begin{equation}
    T_\STOP=E_\RR\bigg(\frac{E_\LL}{T_\LL} +\frac{\Delta E}{T_\HH}\bigg)^{-1}.
\end{equation}
The cooling power drops to zero at $T_\RR=T_\STOP$, as demonstrated in Fig.~\ref{figmm}(a). This hints at the fact that the transport of electrons is thermodynamically reversible at the stopping configuration. This can be straightforwardly verified by showing that Carnot COP is achieved in this situation, see Fig.~\ref{figmm}(b).

\begin{figure}
    \includegraphics[width=\linewidth]{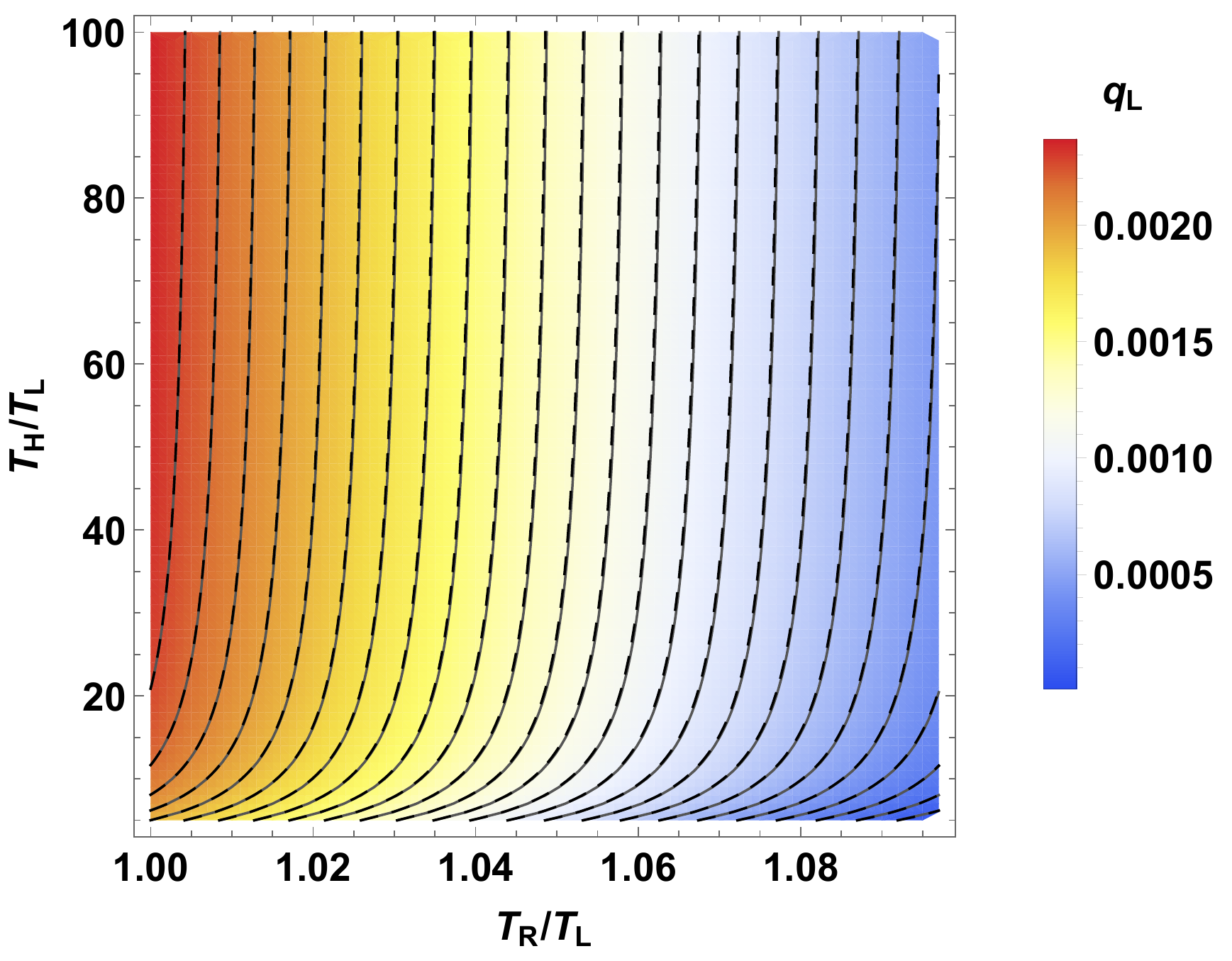}
    \caption{The cooling power~$Q_\LL$ in dimensionless units $q_\LL=\hbar Q_\LL/(\gamma\kB T_\LL)$ as a function of different temperatures involved. The contour lines are obtained independently in the small-linewidth limit from exact calculations (solid lines) and linear response results (dashed lines).  We choose $\mu_\LL=\mu_\RR=0$, $E_\LL=0.4\kB T_\LL$, $E_\RR=0.45\kB T_\LL$. The maximum value of the temperature~$T_\RR$ is taken to be the stopping temperature~$T_\STOP$ calculated when the temperature takes its minimum value~$T_\HH=5T_\LL$.}
    \label{fig4}
\end{figure}
 \section{Optimizing the cooling power}

In this section, we discuss systematically optimizing the performance of our absorption refrigerator. We focus on the parameters which can be tuned via external control, such as the energy of the dots, as well as their linewidth. We compute their optimal values numerically and compare to exact analytical predictions whenever possible.

\subsection{Optimizing w.r.t. the dot energies}

We first discuss the optimal and stopping configurations w.r.t varying the energy of the left dot when all other parameters are assumed to have fixed values. We further assume that we are in the limit of vanishing level width, $\gamma\ll\kB T_\LL$. Hereafter, we assume $j>0$, such that $E_\RR>E_\LL>0$ according to Eq.~\eqref{2ndLaw_v}. In this regime, the second law implies
\begin{equation}
 E_\LL\bigg(\frac{1}{T_\LL}-\frac{1}{T_\HH}\bigg)<E_\RR\bigg(\frac{1}{T_\RR}-\frac{1}{T_\HH}\bigg).
\end{equation} 
We deduce that the system operates as an absorption refrigerator if the dot energies satisfy,
\begin{equation}
     0<E_\LL<r_{\STOP}E_\RR,
     \end{equation}
 where we have introduced the stopping ratio
 \begin{equation}
r_{\STOP}=\frac{T_\RR^{-1}-T_\HH^{-1}}{T_\LL^{-1}-T_\HH^{-1}}.     
 \end{equation}
 The cooling power goes to zero at this stopping configuration where electron transport is thermodynamically reversible and thus achieves Carnot COP. This is demonstrated in Fig.~\ref{figmm}(c) and  Fig.~\ref{figmm}(d) respectively.

We now discuss the optimal point of operation of the refrigerator by first solving for the chemical potential of the reservoir $\mu_\HH$ (see App.~A), and then optimizing the cooling power $Q_\LL$ w.r.t the energy of the left dot. However, the solution for $\mu_\HH$ does not allow for further analytical calculations in the general case and additional approximations are necessary. In what follows, we will assume that the energy differences between dot energies and chemical potentials are small, so that the Fermi factors can be expanded as follows,
\begin{equation}
    f(E-\mu,T)\approx\frac12-\frac{E-\mu}{4\kB T}.
\end{equation}
Such simplification is accurate for $|E-\mu|\ll\kB T$.\footnote{In practice, the results obtained using this approximation hold for much higher energies, typically $|E-\mu|\sim\kB T$.} In this regime, the cavity chemical potential is given by
\begin{equation}
    \mu_\HH\approx-\frac{E_\LL}2\bigg(\frac{T_\HH}{T_\LL}-1\bigg)-\frac{E_\RR}2\bigg(\frac{T_\HH}{T_\RR}-1\bigg).
\end{equation}
The cooling power then reads
\begin{equation}
Q_\LL=\frac{\gamma E_\LL}{8\hbar\kB T_\HH}\Bigg(E_\RR\bigg(\frac{T_\HH}{T_\RR}-1\bigg)-E_\LL\bigg(\frac{T_\HH}{T_\LL}-1\bigg)\Bigg).\label{Q_LR}
\end{equation}
We find that cooling power is maximum when the left dot energy is precisely at the center of its allowed range, $E_\LL^{\mathrm{max}}=r_{\STOP}E_\RR/2$, where
 \begin{equation}
     Q_\LL^{\mathrm{max}}=\frac{\gamma E_\RR^{2}(T_\RR^{-1}-T_\HH^{-1})^{2}}{32\hbar \kB(T_\LL^{-1}-T_\HH^{-1})}.
 \end{equation}
The corresponding COP is
\begin{equation}
    C^\mathrm{max}=\frac{T_\RR^{-1}-T_\HH^{-1}}{2T_\LL^{-1}-T_\RR^{-1}-T_\HH^{-1}}.
\end{equation}
 We indicate these optimal configurations in Fig.~\ref{figmm}(c) and Fig.~\ref{figmm}(d). Further optimization of the device by varying more than one parameter at once are shown in Fig.~\ref{fig3} and Fig.~\ref{fig4}.
 
 Analogous to the rectification of a current without any voltage difference~\cite{jordan2013powerful},  we note from Eq.~\eqref{Q_LR} that it is possible to drive a rectified heat current even if there is no thermal bias between reservoirs~$\LL$ and~$\RR$, $T_\LL=T_\RR=T$. In this case the heat current out of reservoir~$\LL$ is given by
 \begin{equation}
     Q_\LL=\frac{\gamma E_\LL\Delta E}{8\hbar\kB T_\HH}\bigg(\frac{T_\HH}T-1\bigg).
 \end{equation}
 In this situation, the direction of the heat flow, along with the direction of the electric current, is then determined by the relative position of the dot energies. For example, let us consider a setup with $E_\LL>0$ and $E_\RR>0$, in which case heat transport is mediated by hot electrons; we have $Q_\LL>0$ ($j>0$) if $E_\RR>E_\LL$, while $Q_\RR>0$ ($j<0$) if $E_\RR<E_\LL$. More details about this rectified heat current are given in App.~\ref{appB}.
 
\subsection{Optimizing w.r.t. the linewidth $\gamma$}

\begin{figure}
    \includegraphics[width=\linewidth]{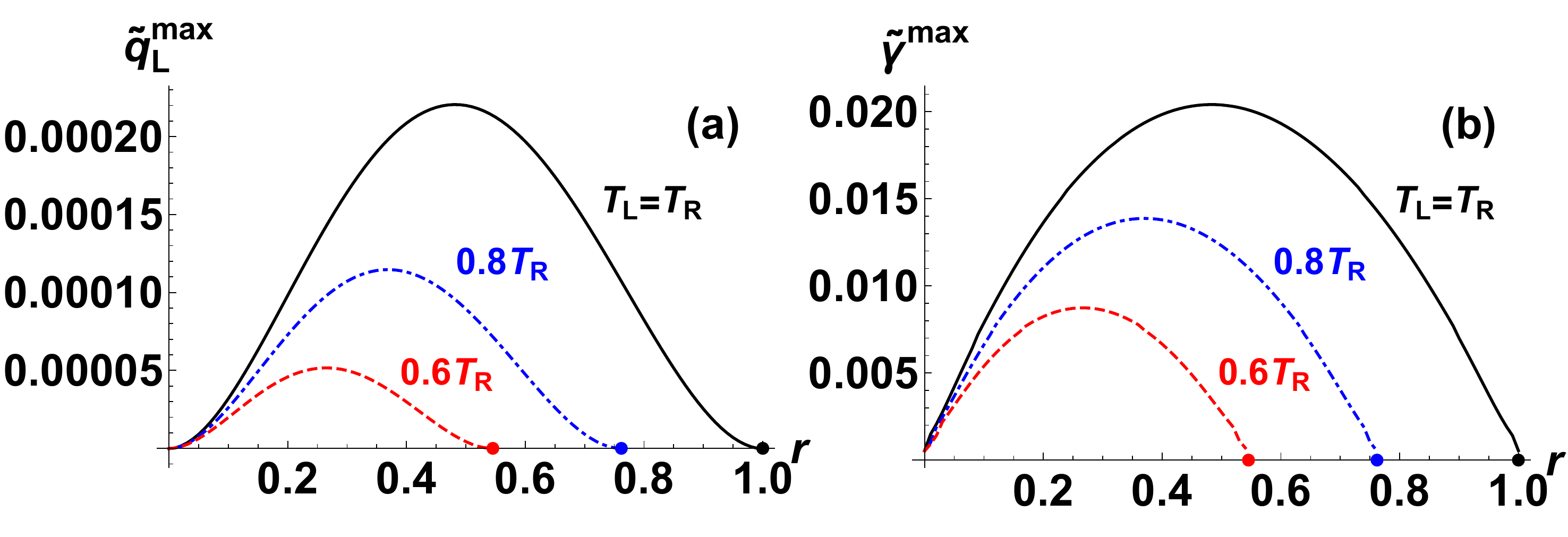}
    \caption{Optimization over finite linewidth. We perform exact numerical simulation of the integrals involved where the energy of the dots and temperatures are not restricted to the regime of validity of linear response theory. (a) The maximum cooling power $Q^{\mathrm{max}}_\LL$ in dimensionless units, $\tilde{q}^{\mathrm{max}}_\LL=\hbar Q^{\mathrm{max}}_\LL/(\kB T_\RR)^{2}$ as a function of the energy ratio $r$ for different value of $T_\LL$.  We choose $\mu_\LL=\mu_\RR=0$, $T_\HH=5T_\RR$, $E_\RR=\kB T_\RR$, and $\gamma=\gamma^{\mathrm{max}}(rE_\RR,E_\RR, T_\LL,T_\RR,T_\HH)$. (b) The maximum cooling power linewidth $\gamma=\gamma^{\mathrm{max}}(rE_\RR,E_\RR, T_\LL,T_\RR,T_\HH)$ in dimensionless units ($\tilde{\gamma}^{\mathrm{max}}=\gamma^{\mathrm{max}}/\kB T_\RR$)  is shown as a function of $r$. The stopping values of $r$ are shown as dots, and they agree well with the predictions from assuming vanishing linewidth. The $T_\LL=T_\RR$ case corresponds to the rectification configuration discussed in App.~\ref{appB}.}
    \label{fig5}
\end{figure}

We now generalize our discussion to finite linewidth $\gamma$ characterizing the transmission through the dots. The linewidth $\gamma$ is an additional important parameter which can be optimized to our advantage in a realistic experiment.  Physically, as we increase the linewidth, electrons from a wider range of energy can participate in the transport process. But soon the cooling power tends to drop with further increase in the linewidth because increasing the linewidth above a threshold reduces the energy filtering effect necessary for the operation of our absorption refrigeration scheme. Therefore there is an optimal linewidth $\gamma=\gamma^{\mathrm{max}}$ which will depend on the energy of the dots as well as the temperature of the leads. We compute $\gamma^{\mathrm{max}}$ numerically. We demonstrate this optimization in Fig.~\ref{fig5}(a), where each point in the curve corresponds to an optimization over the linewidth $\gamma$. The optimal $\gamma$ which maximizes the cooling power for each value of $r$ is shown in Fig.~\ref{fig5}(b). We observe that the allowed range of values for the left dot energies shrinks (from above and below) as the level width is increased. Interestingly, there is a critical level width above which refrigeration becomes impossible. Our numerical results further substantiate our choice to work in the regime of vanishingly small level width since we find that~$\gamma^\mathrm{max}$ is typically  one or two orders of magnitude smaller than the temperature. This is in stark contrast with Ref.~\cite{jordan2013powerful} where the same device is used as an energy harvester. There, the optimal level width for electric power generation is found to be of the order of the temperature of the leads. This substantial difference is seemingly due to the different positions of dot energies: In Ref.~\cite{jordan2013powerful}, dot energies are symmetrically placed with respect to the average chemical potential of the leads, while here, we have argued that refrigeration can only be achieved if both dot energies are above (or below) the common chemical potential of the leads (taken as the zero of energy throughout our analysis).

Numerically analyzing the particle and energy currents as functions of the linewidth~$\gamma$, we find that their large-scale variations with $\gamma$ do not strongly depend on the dot energies. As already noted in Ref.~\cite{jordan2013powerful}, the particle current\footnote{Ref.~\cite{jordan2013powerful} actually focuses on the electric power delivered by the device which is proportional to the particle current analyzed here.} first increases with $\gamma$, reaching a maximum for $\gamma\sim\kB T_\LL$, but it then decreases, approaching zero as $\gamma$ becomes larger. The heat current typically decreases with $\gamma$ and plateaus at $Q_\LL=-\pi^2\kB^2(T_\HH^2-T_\LL^2)/(3h)$ for relatively large linewidths, $\gamma\gtrsim10\kB T_\LL$. More details about the behaviour of currents for large $\gamma$ are given in App.~\ref{appC}. 
Our numerics indicates we must restrict to cases where $\gamma\ll k_{B}T_\LL$.
More precisely, we find that the only cases where the device can operate as a refrigerator are those where the heat current first increases with $\gamma$, but then decreases after having reached a maximum for $\gamma=\gamma^\mathrm{max}$. 
The possibility to use the device as a refrigerator will then be determined by its behaviour for small $\gamma$ which is obtained using the limit in Eq.~\eqref{delta} and has been extensively studied in the present work. 
We have found that $Q_\LL\approx\gamma\hbar^{-1}E_\LL(f(E_\LL,T_\LL)-f(E_\LL-\mu_\HH,T_\HH))$ for $\gamma\ll\kB T_\LL$, which means that the possibility for refrigeration at \textit{any} value of $\gamma$ will be determined by the sign of the initial slope~$\hbar^{-1}E_\LL(f(E_\LL,T_\LL)-f(E_\LL-\mu_\HH,T_\HH))$, refrigeration being possible only if it is positive. Interestingly, we note that the sign of this quantity has already been analyzed in Eq.~\eqref{2ndLaw_v}, which simply is restatement of the second law of thermodynamics in the limit $\gamma\ll\kB T_\LL$. The possibility for our device to operate as a refrigerator is thus entirely determined by the fundamental laws of thermodynamics expressed in the limiting case of vanishingly small linewidths.

\subsection{Comparison to experiments}

We now look at experimentally realistic conditions. The simulation we base our discussion on is shown in Fig.~\ref{figmm}(c). The energy of the right dot considered is $E_\RR=\unit{103}{\micro\electronvolt}$, and is kept fixed. We consider the temperature of the left lead kept at $T_\LL=\unit2\kelvin$, temperature of the right lead kept at $T_\RR=\unit3\kelvin$, and the temperature of the cavity kept at $T_\HH=5T_\RR$. For these considerations, the stopping energy of the left dot becomes, $E_\LL^{\STOP}\approx\unit{63.4}{\micro\electronvolt}$, and the optimal cooing is obtained when $E_\LL^{\mathrm{max}}\approx\unit{31.7}{\micro\electronvolt}$. The maximum cooling power obtained at the optimal bias ($E_\LL=E_\LL^{\mathrm{max}}$) is $Q_\LL^{\mathrm{max}}\approx\unit{10}{\electronvolt\per\second}$. We assume a linewidth $\gamma = 10^{-4}E_\RR$, which approximates the results assuming a delta function linewidth. 

\section{Conclusions}

We presented a new quantum absorption refrigerator scheme based on the quantum physics of resonant tunneling through quantum dots. We provided a complete thermodynamic characterization of the device, identified stopping configurations of the refrigerator where the transport is thermodynamically reversible such that Carnot coefficient of performance is achieved while extracting zero cooling power. We also optimized the operation of the refrigerator w.r.t. to externally controllable parameters for the fridge, such as the energy of the dots, as well as their linewidth. Our absorption refrigerator can be   integrated into circuits, and can offer on-chip integrable solutions to the increasing demand for cooling in the sub-kelvin regime, by harvesting energy from dissipating energy sources within a circuit. This is an additional benefit for our cooling scheme, which presents itself as a novel approach to recycle wasteful energy from some part of the circuit, possibly left over from a cycle of computation, for cooling other regions within the circuit. In contrast to the heat engine mode~\cite{jordan2013powerful}, the absorption refrigerator requires smaller linewidth. Different alternate implementations are possible, for instance, one can consider several of such absorption refrigerators operating in parallel to amplify the cooling effect.  Such practical solutions to harvesting dissipated heat in electronic circuits are of supreme importance to various quantum computing platforms currently available, with the potential to improve the performance of superconducting circuits, quantum limited detectors, and charge sensors  used for various quantum information processing applications.

\section{Acknowledgements} This work is supported by the US Department of Energy (DOE), Office of Science, Basic Energy Sciences (BES), under Grant No. DE-SC0017890.

\appendix 

\begin{widetext}

\section{Exact solution for $\mu_\HH$\label{appA}}

Note that Eq.~\eqref{cc} can be solved exactly to obtain a solution to the chemical potential of the cavity $\mu_\HH$. To find the solution, we define $z=\exp(-\mu_\HH/\kB T_\HH)$, and use the short notation $f_\alpha=f(E_\alpha-\mu_\alpha,T_\alpha)$ for $\alpha=\LL,\RR$. Then, Eq.~\eqref{cc} becomes
\begin{equation}
    \bigg(z \e^{\frac{E_\LL}{\kB T_\HH}}+1\bigg)^{-1}+\bigg(z \e^{\frac{E_\RR}{\kB T_\HH}}+1\bigg)^{-1}=f_\LL+f_\RR\implies\frac{z \e^{\frac{E_\LL}{\kB T_\HH}}+z \e^{\frac{E_\RR}{\kB T_\HH}}+2}{\Big(z \e^{\frac{E_\LL}{\kB T_\HH}}+1\Big)\Big(z \e^{\frac{E_\RR}{\kB T_\HH}}+1\Big)}=f_\LL+f_\RR.
\end{equation}
It is straightforward to write this equation in the form $az^{2}+bz+c=0$, where we find
\begin{equation}
    a=f_\LL+f_\RR,\quad b=\big(f_\LL+f_\RR-1\big)\Big(\e^{-\frac{E_\LL}{\kB T_\HH}}+\e^{-\frac{E_\RR}{\kB T_\HH}}\Big),\quad c=(f_\LL+f_\RR-2)\e^{-\frac{E_\LL+E_\RR}{\kB T_\HH}}.
\end{equation}
There are two solutions to this equation given by $z_{\pm}=(-b\pm\sqrt{b^{2}-4ac})/(2a)$. Note that we always have $b^2>4ac$ since $a\geq 0$ and $c\leq0$, which ensures that the solutions~$z_\pm$ are real. However, this also implies that $z_-\leq0$ which therefore is an unphysical solution as it would correspond to an imaginary chemical potential. We conclude that the chemical potential of the cavity reads
\begin{equation}
    \frac{\mu_\HH}{\kB T_\HH}=\ln2-\ln\Bigg[\begin{aligned}[t]
    &\Big(\frac1{f_\LL+f_\RR}-1\Big)\Big(\e^{-\frac{E_\LL}{\kB T_\HH}}+\e^{-\frac{E_\RR}{\kB T_\HH}}\Big)\\
    &+\bigg(\Big(\frac1{f_\LL+f_\RR}-1\Big)^2\Big(\e^{-\frac{E_\LL}{\kB T_\HH}}+\e^{-\frac{E_\RR}{\kB T_\HH}}\Big)^2+4\Big(\frac2{f_\LL+f_\RR}-1\Big)\e^{-\frac{E_\LL+E_\RR}{\kB T_\HH}}\bigg)^{\!\!\!1/2}\Bigg].
    \end{aligned}
\end{equation}
We emphasize that the above expression can be used in all situations where the narrow-linewidth limit is justified since its derivation did not require any additional approximation. In particular, even though this work focuses on the absorption refrigerator case where $\mu_\LL=\mu_\RR=0$, the chemical potentials~$\mu_\LL$ and~$\mu_\RR$ need not be equal.

 \section{Rectification configuration\label{appB}}
 
It is interesting to note that it is possible to drive a rectified heat current even if there is no thermal bias between the two colder reservoirs, that is $T_\LL=T_\RR=T_\CC$. In this case, the second law of thermodynamics reads
\begin{equation}
    J\bigg(\frac1{T_\CC}-\frac1{T_\HH}\bigg)\ge0.
\end{equation}
Since $T_\CC<T_\HH$, this imposes $J>0$, that is $j\Delta E>0$. In this case the directions of both the heat flow and the electric current are determined by the relative position of the dot energies; when $E_\LL>0$ and $E_\RR>0$, we have $Q_\LL>0$ ($j>0$) provided $E_\RR>E_\LL$, while $Q_\RR>0$ ($j<0$) provided $E_\RR<E_\LL$.

Focusing on the case~$j>0$, we realize that the only condition to be satisfied by the dot energies is $\Delta E>0$; in other words, $r_\STOP=1$. In the linear response regime, we find that the cooling power is given by
\begin{equation}
    Q_\LL=\frac{\gamma E_\LL\Delta E}{8\hbar k_\mathrm BT_\HH}\bigg(\frac{T_\HH}{T_\CC}-1\bigg).
\end{equation}
For fixed $E_\RR$, it reaches a maximum when $E_\LL=E_\RR/2$,
\begin{equation}
    Q_\mathrm{max}=\frac{\gamma E_\RR^2}{32\hbar k_\mathrm BT_\HH}\bigg(\frac{T_\HH}{T_\CC}-1\bigg),
\end{equation}
 with the coefficient of performance simply given by $C=1$.
 
 \section{Large-linewidth limit: flat transmission\label{appC}}

The transmission function of a quantum dot becomes flat in the limit of large linewidth,
\begin{equation}
    \mathcal T_\alpha(E)=\frac{\gamma^2}{(E-E_\alpha)^2+\gamma^2}\approx1.
\end{equation}
In practice, such approximation is relevant when $\gamma$ is much larger than temperature, $\gamma\gg\kB T_\HH$ here. In this regime, the particle currents read
\begin{equation}
    j_\alpha\approx\frac2h\int_{-\infty}^{\infty}\mathrm dE\,(f(E-\mu_\alpha,T_\alpha)-f(E-\mu_\HH,T_\HH))=\frac2h(\mu_\alpha-\mu_\HH).
\end{equation}
The conservation law in Eq.\eqref{cc} then becomes
\begin{equation}
    j_\LL+j_\RR=0\implies\mu_\HH=\frac{\mu_\LL+\mu_\RR}2.
    \label{muH_inf}
\end{equation}
Hence, the particle current going through the device is
\begin{equation}
    j=j_\LL=-j_\RR=\frac{\mu_\LL-\mu_\RR}h.
\end{equation}
Furthermore, we can compute the energy currents,
\begin{equation}
    J_\alpha\approx\frac2h\int_{-\infty}^{\infty}\mathrm dE\,E(f(E-\mu_\alpha,T_\alpha)-f(E-\mu_\HH,T_\HH))=-\frac{\pi^2\kB^2}{3h}(T_\HH^2-T_\alpha^2)+\frac1h(\mu_\alpha^2-\mu_\HH^2).
\end{equation}
Using the expression for $\mu_\HH$ in Eq.~\eqref{muH_inf}, we obtain
\begin{equation}
J_\alpha=-\frac{\pi^2\kB^2}{3h}(T_\HH^2-T_{\alpha}^2)+\frac{\mu_\alpha-\mu_{\bar\alpha}}{4h}(3\mu_\alpha+\mu_{\bar\alpha}),
\end{equation}
where $\bar\alpha=\RR$ if $\alpha=\LL$, and conversely.

In this article, we have focused on the case of an absorption refrigerator where $\mu_\LL=\mu_\RR=0$. In such a situation, we find $\mu_\HH=0$ for $\gamma\gg\kB T_\HH$. This implies that the particle current vanishes in this limit, $j=0$. In contrast, the energy current remain finite,
\begin{equation}
    J_\alpha=-\frac{\pi^2\kB^2}{3h}(T_\HH^2-T_\alpha^2).
\end{equation}
In particular, the heat current out of the cold reservoir~$\LL$ is given by $Q_\LL=J_\LL=-\pi^2\kB^2(T_\HH^2-T_\LL^2)/(3h)$. Interestingly, this limiting value does not depend on the dot energies. This is natural since the Lorentzian resonances at the dots' energies are completely blurred out in the large-linewidth limit where the transmission functions become flat. Moreover, we find that this heat current is always negative which further substantiates our claim that refrigeration is only possible for small linewidths.

  \end{widetext}
  
\bibliography{Reference.bib}
\bibliographystyle{ieeetr}
		   \end{document}